\documentclass[aps,prb,reprint]{revtex4-1}
\usepackage{graphicx}
\usepackage{color}
\usepackage{amsmath}

\begin{document}
	
\title{First-principles study of magnetism, lattice dynamics, and superconductivity in LaFeSiH$_x$}
\author{Linda Hung}
\altaffiliation[Present address: ]{Toyota Research Institute, 4440 El Camino Real, Los Altos, California 94022, USA}
\affiliation{NIST Center for Neutron Research, National Institute of Standards and Technology, Gaithersburg, Maryland 20899-6102, USA}
\author{Taner Yildirim}
\affiliation{NIST Center for Neutron Research, National Institute of Standards and Technology, Gaithersburg, Maryland 20899-6102, USA}
\email{taner@nist.gov}
\date{\today}
	
\begin{abstract}
The structural, electronic, magnetic, and vibrational properties of LaFeSiH$_x$ for $x$ between 0 and 1 are investigated using
density functional calculations. We find that the electronic and magnetic properties are strongly controlled by the hydrogen concentration $x$ in  LaFeSiH$_x$.  While fully hydrogenated LaFeSiH has a striped antiferromagnetic ground state, the underdoped LaFeSiH$_x$ for $x\leq0.75$ is not magnetic within the virtual crystal approximation or with explicit doping of supercells.   The antiferromagnetic configuration breaks the symmetry of Fe $d$ orbitals and increases electron-phonon coupling up to $50\%$, especially for modes in the 20-50 meV range that are associated with Fe atomic movement. We find competing nearest and next-nearest neighbor exchange interactions and significant spin-phonon coupling, qualitatively similar but smaller in magnitude compared those found in LaOFeAs superconductors. The superconducting $T_c$ for antiferromagnetic LaFeSiH$_x$, assuming conventional superconductivity via McMillan's equation, therefore is computed to be 2-10 K, in contrast to $T_c\approx0$ for the nonmagnetic material. We also predict that the  LaFeSiH$_x$
could be a good proton conductor due to phase stability with a wide range of hydrogen concentration $x < 1$. 
\end{abstract}
	
\keywords{superconductivity, phonons}
\maketitle

\section{Introduction}

Iron-based superconductors have been widely studied since the discovery of fluorine-doped LaFeAsO with superconductivity near 26 K.\cite{Kamihara2008,Paglione2010,*Johnston2010,*Stewart2011,*Hosono2015}  This class of materials are metals in the normal state and have been found to superconduct upon doping parent compounds that show antiferromagnetic phase transitions. Hence, it is widely believed that the strong antiferromagnetic spin fluctuations are responsible for the observed high-$T_c$ superconductivity.

Iron-based superconductors are layered with a square lattice configuration, where the Fe atoms lie in a plane and another element -- either chacogenides or pnictogens -- are slightly above and below.  Due to toxicity of pnictogens/chacogenides, it is of interest to discover new analogous Fe-based materials without pnictogens/chacogenides.

Recently, a new iron-containing material, LaFeSiH, has been reported to show evidence of superconductivity $\sim$8 K.\cite{Bernardini2017}  As with other Fe-based superconductors, the room-temperature structure at 293 K has a tetragonal P4/nmm symmetry (space group 129), while the low-temperature nematic state at 15 K is an orthorhombic Cmma (s.g. 67).  However, this material, composed of alternating layers of La-H and Fe-Si, is chalcogenide- and pnictogen-free. It is therefore of interest to find out the similarities and differences between the LaFeSiH$_x$ and the LaFeAsO systems.  Futhermore, we would like to determine how the magnetic and structural properties of LaFeSiH$_x$ depend on the hydrogen concentration $x$, since earlier studies suggest that ternary rare earth transition metal silicides (i.e. TMSi, TM=Co, Mn, etc) have very rich magnetic phase diagram over a wide range of hydrogen concentrations\cite{Tence2009,Tence2010,Welter1998}. 

In this work, we use first-principles methods -- density-functional theory (DFT) and density functional perturbation theory (DFPT) -- to better understand LaFeSiH, examining how the proportion of H affects the magnetic state, how the magnetic state affects the orbital occupations and structure, and how all these factors affect superconductivity.  For our DFT computations, we use Quantum Espresso,\cite{Giannozzi2009} the GBRV ultrasoft pseudopotentials,\cite{Garrity2014} the Perdew-Burke-Ernzerhof exchange-correlation functional,\cite{Perdew1996} $8\times8\times6$ $k$-point sampling, and 0.01 Ry Methfessel-Paxton smearing for simulation cells containing four formula units of LaFeSiH$_x$.

\section{Stoichiometric L\MakeLowercase{a}F\MakeLowercase{e}S\MakeLowercase{i}H}

We optimize the atomic positions and determine the electronic structure of LaFeSiH with nonmagnetic (NM), ferromagnetic (FM), checkerboard antiferromagnetic (cAFM), and single stripe antiferromagnetic (sAFM) starting configurations.  These computations confirm that LaFeSiH has a sAFM ground state, differing from the optimized NM configuration by $<3$~meV per formula unit.  In Table~\ref{tab:structure}, we list the optimized atomic positions associated with lattice parameters from low-temperature (15 K) and room-temperature (293 K) powder diffraction measurements,\cite{Bernardini2017} as well as those using DFT-optimized lattice dimensions.  The sAFM spin configuration results in an orthorhombic distortion of the optimized lattice, indicating strong magneto-elastic interactions in LaFeSiH; this is also a characteristic feature of other Fe-pnictide based superconductors.  The Si in LaFeSiH acts analogously to the As in LaOFeAs, whose position is critically controlled by the magnetic properties of the superconductor\cite{chalcogen_height,yildirim_physica}. However, note that results are not quantitative: the lattice anisotropy is greater than observed in experiment (parameters differ by about 0.1~\AA), and while there is good agreement between the simulated and experimental La positions, the optimized Si atoms are up to 0.2~\AA~away from the measured positions.  
Nevertheless, these similarities are very promising and suggest that 
LaFeSiH$_x$ may also be superconducting at high temperatures after suitable doping by either hydrogenation or by other elements at the La-site.

\begin{table}
	\caption{\label{tab:structure}Fractional atomic positions of LaFeSiH for experimental and DFT lattice dimensions.} 
	\begin{ruledtabular}
		\begin{tabular}{lccccc}
			&$a$	&$b$	&$c$	&La($z$)	&Si($z$)\\
			\hline
			293 K\cite{Bernardini2017} 	&5.6950	&5.6950	&8.0374	&0.6722	&0.1500\\
			NM	&5.6950	&5.6950	&8.0374	&0.6794	&0.1310\\ 
			sAFM	&5.6950	&5.6950	&8.0374	&0.6775	&0.1348\\ 
			\hline
			15 K\cite{Bernardini2017} 	&5.6831	&5.7039	&7.9728	&0.6747	&0.155\\
				NM	&5.6831	&5.7039	&7.9728	&0.6788	&0.1319\\
			sAFM	&5.6831	&5.7039	&7.9728	&0.6768	&0.1357\\
			\hline
			DFT opt. \\
			NM &5.723	&5.723	&7.843	&0.6784	&0.1325\\
			sAFM &5.648	&5.738	&7.932	&0.6759	&0.1375\\
		\end{tabular}
	\end{ruledtabular}
\end{table}

To gain more insight into the magnetic properties of LaFeSiH, we perform constrained magnetization calculations for the FM, cAFM, and sAFM configurations.  Fixing atomic positions to that of the sAFM ground state and using the 15 K lattice parameters, we test for magnetizations ranging from 0.2 $\mu_B$ to 1.0 $\mu_B$ per Fe.  These energy differences, relative to the nonmagnetic state, are also used to determine the contributions of the nearest neighbor interactions ($J_1$) and next-nearest neighbor interactions ($J_2$), as defined within a Heisenberg model, where $H=\sum_{i,j}J_{ij}M_iM_j$ over pairs $ij$.  Note that while LaFeSiH is not fully localized, a microscopic description of the magnetic interactions obtained via $J_1$ and $J_2$ may nonetheless allows additional insight into material properties.\cite{Yildirim2008}  These computations show that $J_1<2J_2$ over range of magnetizations studied, again validating the sAFM ground state. At the optimized magnetization near 0.8 $\mu_B$, both exchange interactions $J_1$ and $J_2$ are of comparable magnitude ($\sim6$ meV).
Even though these magnetic interactions are large, it is significantly less than those found in Fe-pnictide based superconductors\cite{yildirim_physica}.

\begin{figure}
	\includegraphics[scale=0.9]{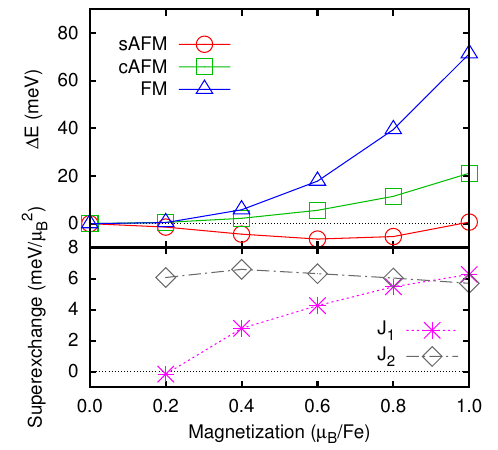}
	\caption{\label{fig:superexchange} Total energy per Fe atom vs. magnetic moment for the FM, sAFM, and cAFM spin configurations, and effective $J_1$ and $J_2$ obtained from their energy differences at various magnitudes of magnetization.}
\end{figure}

\begin{figure}
	\includegraphics[scale=1.0]{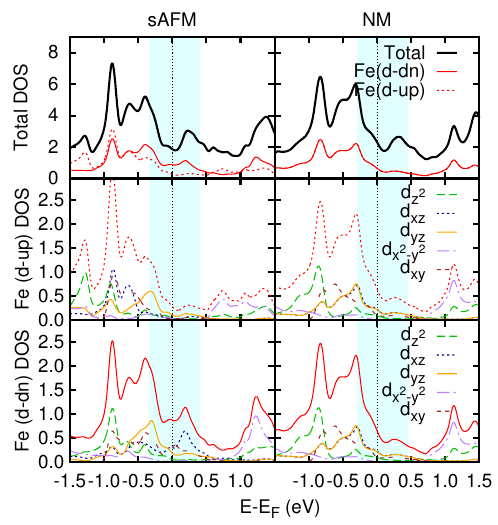}
	\caption{\label{fig:dos_electronic}Electronic DOS per formula unit for sAFM and NM LaFeSiH (15 K lattice parameters), with blue shading indicating the range of Fermi level shifts with up to 1 electron or hole doping and the rigid band approximation.}
\end{figure}

The effect of antiferromagnetism on the electronic DOS is shown in Figure~\ref{fig:dos_electronic}.  For the total antiferromagnetic simulation cell, the DOS associated with each spin channel does not appear to be spin-polarized due to the balanced number of up and down spin Fe atoms.  We therefore show the total and projected DOS for a single formula unit.  For NM LaFeSiH, despite the imposed nematic lattice, the $xz$ and $yz$ orbitals are essentially degenerate and together comprise 40\% of the DOS at the Fermi level; the $xy$ orbitals contribute 9\%.  In the sAFM configuration, $xz$ and $yz$ orbital symmetry is broken, with the 3$d_{xz}$ states shifting to lower energies in one spin channel and to higher energies in the other.  One of the $d$ spin channels (labeled here as ``dn") is the larger contributor to the Fermi level DOS; its $yz$ orbital contributes 10\%, its $xz$ orbital contributes 16\%, and its $xy$ states contribute 17\% to the total DOS (including both spins).  DFT therefore predicts that orbital fluctuations in LaFeSiH can only occur if there are also spin fluctuations. The blue shaded region shown in Figure~\ref{fig:dos_electronic} indicates that, within the rigid band approximation, one can potentially double the density of states at the Fermi level by fractional hole doping. Hence, doping either via divalent metal substitution at the La-site or varying hydrogen concentration $x$ could yield a material with enhanced
electronic/superconducting properties in LaFeSiH$_x$ system.

\begin{figure}
	\includegraphics[scale=1.0]{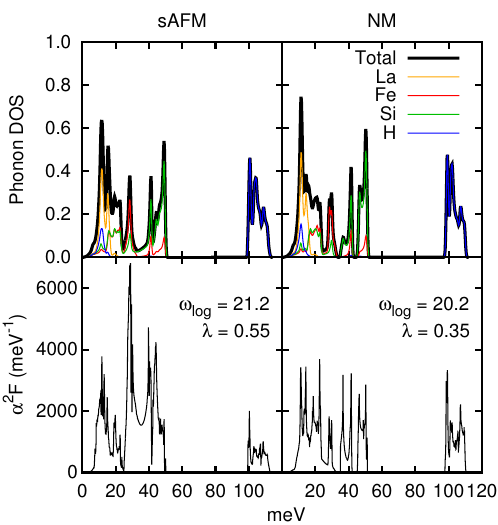}
	\caption{\label{fig:dos_a2F}Phonon DOS and Eliashberg functions of sAFM and NM LaFeSiH (15 K lattice parameters).}
\end{figure}

For the lattice dynamics of LaFeSiH, the effects of magnetism are less dramatic, compared to one found in LaOFeAs 
system\cite{yildirim_physica}. Density functional perturbation theory (DFPT) calculations on a $2x2x2$ $\Gamma$-centered grid confirm the local stability of both NM and sAFM configurations with Cmma symmetry; all phonons are found to have real energies.  The phonon DOS for both NM and sAFM configurations predict La-coupled low-energy modes below 20 meV, Fe and Si modes between 20-50 meV, and high-energy H modes above 100 meV (Fig.~\ref{fig:dos_a2F}).  Gaps in the NM phonon DOS are observed at 33-35 meV and 42-45 meV; these are not observed for the sAFM configuration.

The combined influence of electrons and phonons is evident in plots of the Eliashberg function $\alpha^2F$ (Fig.~\ref{fig:dos_a2F}).  In particular, the Eliashberg function for sAFM LaFeSiH has its largest peaks at 29 meV and 40 meV.  These phonon frequencies are associated with Fe atoms, and the large increase in electron-phonon coupling is consistent with the changes in the electronic structure of Fe atomic orbitals in the magnetic state.

Using $\alpha^2F$, we compute the mass enhancement parameter $\lambda$ and determine the superconducting temperature via McMillan's equation.\cite{McMillan1968}  Using 6 mRy broadening of the electronic states and a 2x2x2 $\Gamma$-centered phonon grid, we compute $\lambda$, $\omega_\mathrm{log}$, and $T_c$ as reported in Table~\ref{tab:structure-Tc}.
We note that the stripe magnetic ordering enhances the electron-phonon coupling almost $60\%$, which is very similar
to the magnetic enhancement of $\lambda$ found other Fe-based superconductors\cite{boeri2010}. Using the limiting value for the Coulomb pseudopotential ($\mu^*=0$), we find that $T_c<10.9$ K for sAFM and $T_c<3.6$ K for NM LaFeSiH; with the commonly-used value of $\mu^*=0.15$, the superconducting transition temperatures are 2.0 K for sAFM and 0.0 K for NM.  Therefore, the experimentally observed transition at 8.5 K, which has been attributed to superconductivity,\cite{Bernardini2017} lies at the high end of the range of predicted superconducting temperatures, after accounting for the magnetic structure and assuming conventional 
electron-phonon mediated superconductivity.

In conclusion for the stoichiometric LaFeSiH system, we find that magnetic interactions, spin-phonon and
electron-phonon couplings, Si-height, etc are all very similar to those found in FeAs-based systems even though
the effects/interactions are smaller in the case of LaFeSiH. 
Based on electronic DOS shown in Fig.~\ref{fig:dos_electronic} and the fact that hydrogen concentration
can be varied in LaFeSiH$_x$\cite{Welter1998}, it may be possible to tune some of the magnetic and electronic properties
with hydrogen concentration and optimize the superconductivity temperature. 
In the next section, we explore this possibility.
 
\section{Effect of H vacancies}

We now simulate the structural, magnetic, and superconductive effects of vacancies in LaFeSiH$_x$, ranging from LaFeSiH$_{1}$ as studied in the previous section, through LaFeSiH$_{0}$, which has P4/nmm symmetry.\cite{Welter1998}  Simulations are performed by explicitly adding or removing H atoms, using simulation cells containing 4 Fe atoms and allowing lattice dimensions to relax.  By symmetry, there is only one unique supercell for (LaFeSiH$_{0.25}$)$_4$, and one for (LaFeSiH$_{0.75}$)$_4$.  On the other hand, there are multiple configurations of the (LaFeSiH$_{0.5}$)$_4$ simulation cell.

\begin{table*}
	\caption{\label{tab:structure-Tc}Structural, magnetic, and superconducting properties of LaFeSiH$_x$ from DFT, where lattice dimensions (optimized unless otherwise denoted by a citation) are in~\AA, $z$ are in fractional coordinates, magnetization is $\mu_B$ per Fe.}
	\begin{ruledtabular}
		\begin{tabular}{lcccccccccccc}
			x&$a$	&$b$	&$c$	&La($z$)	&Si($z$)	&$N_F$	&$\mu_B$	&$N_F$(br.)	&$\lambda$	&$\omega_\mathrm{log}$	&$T_c(\mu^*=0.0)$	&$T_c(\mu^*=0.15)$\\
			\hline
			1.00\footnote[2]{Lattice dimensions from Ref.~\onlinecite{Bernardini2017}} (NM)	&5.6831	&5.6831	&7.9728	&0.6788	&0.1319	&2.43	&0.00	&2.47	&0.35	&20.2	&3.6	&0.0\\
			1.00\footnote[2]{Lattice dimensions from Ref.~\onlinecite{Bernardini2017}}	(sAFM) &5.6831	&5.6831	&7.9728	&0.6768	&0.1357	&1.57	&0.71	&1.89	&0.55	&21.2	&10.9	&2.0\\
			\hline
			1.00	&5.648	&5.738	&7.932	&0.6759	&0.1375	&1.57	&0.90	&1.60	&0.41	&21.4	&5.7	&0.3\\ 
			0.75	&5.743	&5.743	&7.713	&0.6839	&0.1357	&1.76	&0.00	&1.86	&0.29	&19.9	&1.9	&0.0\\
			0.50s	&5.764	&5.756	&7.599	&0.6888	&0.1396	&1.86	&0.00	&1.96	&0.27	&18.0	&1.3	&0.0\\
			0.50c	&5.754	&5.754	&7.534	&0.6866	&0.1417	&1.67	&0.00	&1.70	&0.27	&18.5	&1.4	&0.0\\
			0.25	&5.771	&5.771	&7.423	&0.6911	&0.1461	&2.26	&0.00	&2.31	&0.34	&16.7	&2.7	&0.0\\
			0.00	&5.785	&5.785	&7.219	&0.6917	&0.1537	&2.83	&0.00	&2.82	&0.44	&15.2	&4.9	&0.4\\ 
		\end{tabular}
	\end{ruledtabular}
\end{table*}

\begin{figure}
	\includegraphics[scale=1.0]{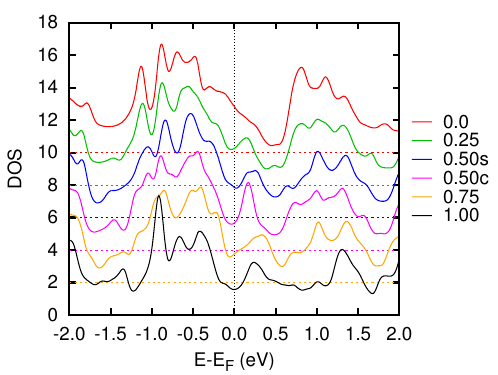}
	\caption{\label{fig:dos-H}Total electronic DOS for LaFeSiH$_x$ with $x$ between 0 and 1, plotted with offsets of 2 (electrons per formula unit per eV).}
\end{figure}

Our computations indicate that at $x=0.0$, 0.25, 0.5, and 0.75, the ground state structure is not magnetic.  There are thus two possible configurations of (LaFeSiH$_{0.5}$)$_4$ supercells, with H vacancies arranged to either form stripes (s) or a checkerboard pattern (c).  In all cases, the LaH and FeSi layers all remain sharply defined in the $z$ direction (Table~\ref{tab:structure-Tc}), with the in-plane coordinates for optimized atomic positions lying at $(-\delta_x,0.25+\delta_y)$, $(-\delta_x,0.75-\delta_y)$, $(0.5+\delta_x,0.25+\delta_y)$, and $(0.5+\delta_x,0.75-\delta_y)$, where $\delta_x$ and $\delta_y$ depend on the layer and specific pattern of vacancies.  Because of the rearrangements in unit cell geometry, the overall electronic DOS line shapes are qualitatively dissimilar beyond a simple potential shift when varying $x$ in LaFeSiH$_x$ (Figure~\ref{fig:dos-H}). Our calculations clearly indicate that the hydrogenation of LaFeSi has a huge effect on the magnetic properties of the resulting compound LaFeSiH$_x$, a similar situation to that observed in the hydrogenation of the other ternary silicide such as NdMnSi\cite{Tence2009}.

\begin{figure}
	\includegraphics[scale=1.0]{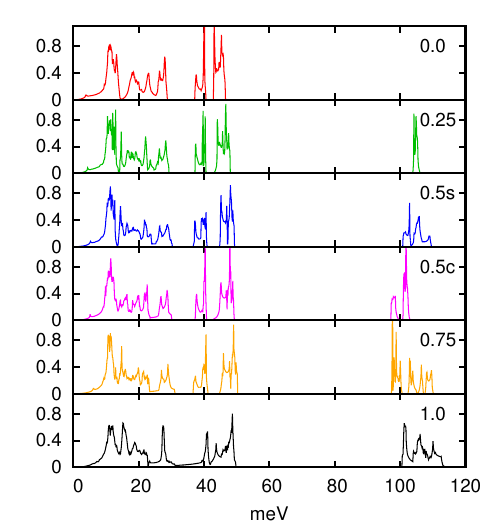}
	\caption{\label{fig:pdos-H}Phonon DOS for LaFeSiH$_x$ with $x$ between 0 and 1.}
\end{figure}

From DFPT, we confirm that these structures with fractional H occupation are all stable.  The phonon DOS changes primarily above 100 meV, consistent with the changing H fraction (Fig.~\ref{fig:pdos-H}).  In addition, as long as the ground state remains nonmagnetic, the superconductive $T_c$ associated with McMillan's equation remains close to zero (Table~\ref{tab:structure-Tc}).  The predicted temperature becomes larger only if the normal state is sAFM.

\begin{figure}
	\includegraphics[scale=1.0]{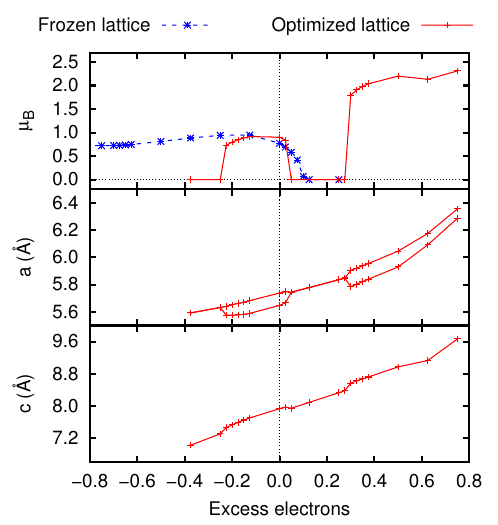}
	\caption{\label{fig:virtualdoping}Effect on magnetization due to virtual doping (per Fe), either with a frozen lattice or with lattice parameters allowed to relax.}
\end{figure}

Large supercells would be needed to explicitly sample fractional occupations of H with finer resolution.  We instead finally simulate fractional occupations in LaFeSiH$_x$ using ``virtual" doping, in which (some fraction of) an electron or hole is added to the system, and offset by a uniform background charge.  Taking LaFeSiH as the parent material, we increment doping levels by as little as 0.025 electrons per formula unit, optimize the lattice parameters, and observe that sAFM structure first appears at 0.225 hole doping (Fig.~\ref{fig:virtualdoping}).  LaFeSiH becomes NM at 0.05 electron doping, but interestingly again transitions to sAFM at 0.3 electron doping.

We contrast the above results with the same computations made with lattice parameters fixed at the 15 K experimental values.  With a frozen lattice, the phase diagram is simplified: electron doping results in the loss of magnetization at approximately 0.1 excess electrons per Fe, while hole doping maintains the sAFM spin configuration.  Therefore, the phase diagram of LaFeSiH$_x$ is shown to arise from the complex interaction of chemical potential together with the chemical pressure.  These trends associated with doping parallel the predicted disappearance and reappearance of magnetization for LaFeSiH under pressure.\cite{Bernardini2017}  Therefore, the quantitative amount of doping required to induce phase changes should be very sensitive to both the choice of dopant and external strain on the material.

\section{Conclusions}
We investigate properties of LaFeSiH$_x$, particularly in relation to the possible superconductive phase transition near 8 K reported in recent work.\cite{Bernardini2017}  The sAFM ground state of LaFeSiH affects its electronic structure by breaking the symmetry of the $d_{xz}$ and $d_{yz}$ orbitals.  Phonon modes are characterized; the intermediate modes ranging from 20-50 meV are associated with the movement of Fe atoms and exhibit increased electron-phonon coupling for sAFM LaFeSiH, compared to the nonmagnetic material.

First-principles calculations indicate that LaFeSiH$_x$ with fractional $x$ is stable and has well-defined layers, with interlayer spacing increasing with $x$.  LaFeSiH$_x$ with $x$ at 0.75, 0.50, 0.25, and 0.0 have nonmagnetic ground states and are predicted via McMillan's equation to have very low superconducting temperatures.  Our results indicate that the parent material must be sAFM to achieve superconductivity near 8 K.  While the phase diagram of LaFeSiH is extremely sensitive to the choice of dopant, small amounts of doping in LaFeSiH may be able to maintain its sAFM spin configuration, while altering the chemical potential and chemical pressure enough to tune the value of $T_c$.
It would be interesting to confirm the sensitivity of magnetism on hydrogen concentration predicted in this study.
Finally, we also note that LaFeSiH$_x$ with fractional H occupation may be a good proton conductor due to the
fast diffusion of protons in the solid at high temperatures.   However, this is outside of the scope of this work and will be discussed elsewhere.

\section{Acknowledgements}
L.H. acknowledges support from the NIST Director's Postdoctoral Fellow Program.

\bibliography{LaFeSiH}

\begin{thebibliography}{17}%
\makeatletter
\providecommand \@ifxundefined [1]{%
 \@ifx{#1\undefined}
}%
\providecommand \@ifnum [1]{%
 \ifnum #1\expandafter \@firstoftwo
 \else \expandafter \@secondoftwo
 \fi
}%
\providecommand \@ifx [1]{%
 \ifx #1\expandafter \@firstoftwo
 \else \expandafter \@secondoftwo
 \fi
}%
\providecommand \natexlab [1]{#1}%
\providecommand \enquote  [1]{``#1''}%
\providecommand \bibnamefont  [1]{#1}%
\providecommand \bibfnamefont [1]{#1}%
\providecommand \citenamefont [1]{#1}%
\providecommand \href@noop [0]{\@secondoftwo}%
\providecommand \href [0]{\begingroup \@sanitize@url \@href}%
\providecommand \@href[1]{\@@startlink{#1}\@@href}%
\providecommand \@@href[1]{\endgroup#1\@@endlink}%
\providecommand \@sanitize@url [0]{\catcode `\\12\catcode `\$12\catcode
  `\&12\catcode `\#12\catcode `\^12\catcode `\_12\catcode `\%12\relax}%
\providecommand \@@startlink[1]{}%
\providecommand \@@endlink[0]{}%
\providecommand \url  [0]{\begingroup\@sanitize@url \@url }%
\providecommand \@url [1]{\endgroup\@href {#1}{\urlprefix }}%
\providecommand \urlprefix  [0]{URL }%
\providecommand \Eprint [0]{\href }%
\providecommand \doibase [0]{http://dx.doi.org/}%
\providecommand \selectlanguage [0]{\@gobble}%
\providecommand \bibinfo  [0]{\@secondoftwo}%
\providecommand \bibfield  [0]{\@secondoftwo}%
\providecommand \translation [1]{[#1]}%
\providecommand \BibitemOpen [0]{}%
\providecommand \bibitemStop [0]{}%
\providecommand \bibitemNoStop [0]{.\EOS\space}%
\providecommand \EOS [0]{\spacefactor3000\relax}%
\providecommand \BibitemShut  [1]{\csname bibitem#1\endcsname}%
\let\auto@bib@innerbib\@empty
\bibitem [{\citenamefont {Kamihara}\ \emph {et~al.}(2008)\citenamefont
  {Kamihara}, \citenamefont {Watanabe}, \citenamefont {Hirano},\ and\
  \citenamefont {Hosono}}]{Kamihara2008}%
  \BibitemOpen
  \bibfield  {author} {\bibinfo {author} {\bibfnamefont {Y.}~\bibnamefont
  {Kamihara}}, \bibinfo {author} {\bibfnamefont {T.}~\bibnamefont {Watanabe}},
  \bibinfo {author} {\bibfnamefont {M.}~\bibnamefont {Hirano}}, \ and\ \bibinfo
  {author} {\bibfnamefont {H.}~\bibnamefont {Hosono}},\ }\href {\doibase
  10.1021/ja800073m} {\bibfield  {journal} {\bibinfo  {journal} {J. Am. Chem.
  Soc.}\ }\textbf {\bibinfo {volume} {130}},\ \bibinfo {pages} {3296} (\bibinfo
  {year} {2008})}\BibitemShut {NoStop}%
\bibitem [{\citenamefont {Paglione}\ and\ \citenamefont
  {Greene}(2010)}]{Paglione2010}%
  \BibitemOpen
  \bibfield  {author} {\bibinfo {author} {\bibfnamefont {J.}~\bibnamefont
  {Paglione}}\ and\ \bibinfo {author} {\bibfnamefont {R.~L.}\ \bibnamefont
  {Greene}},\ }\href {\doibase 10.1038/nphys1759} {\bibfield  {journal}
  {\bibinfo  {journal} {Nat Phys}\ }\textbf {\bibinfo {volume} {6}},\ \bibinfo
  {pages} {645} (\bibinfo {year} {2010})}\BibitemShut {NoStop}%
\bibitem [{\citenamefont {Johnston}(2010)}]{Johnston2010}%
  \BibitemOpen
  \bibfield  {author} {\bibinfo {author} {\bibfnamefont {D.~C.}\ \bibnamefont
  {Johnston}},\ }\href {\doibase 10.1080/00018732.2010.513480} {\bibfield
  {journal} {\bibinfo  {journal} {Advances in Physics}\ }\textbf {\bibinfo
  {volume} {59}},\ \bibinfo {pages} {803} (\bibinfo {year} {2010})}\BibitemShut
  {NoStop}%
\bibitem [{\citenamefont {Stewart}(2011)}]{Stewart2011}%
  \BibitemOpen
  \bibfield  {author} {\bibinfo {author} {\bibfnamefont {G.~R.}\ \bibnamefont
  {Stewart}},\ }\href {\doibase 10.1103/RevModPhys.83.1589} {\bibfield
  {journal} {\bibinfo  {journal} {Rev. Mod. Phys.}\ }\textbf {\bibinfo {volume}
  {83}},\ \bibinfo {pages} {1589} (\bibinfo {year} {2011})}\BibitemShut
  {NoStop}%
\bibitem [{\citenamefont {Hosono}\ and\ \citenamefont
  {Kuroki}(2015)}]{Hosono2015}%
  \BibitemOpen
  \bibfield  {author} {\bibinfo {author} {\bibfnamefont {H.}~\bibnamefont
  {Hosono}}\ and\ \bibinfo {author} {\bibfnamefont {K.}~\bibnamefont
  {Kuroki}},\ }\href {\doibase 10.1016/j.physc.2015.02.020} {\bibfield
  {journal} {\bibinfo  {journal} {Physica C: Superconductivity and its
  Applications}\ }\bibinfo {series} {Superconducting {Materials}:
  {Conventional}, {Unconventional} and {Undetermined}},\ \textbf {\bibinfo
  {volume} {514}},\ \bibinfo {pages} {399} (\bibinfo {year}
  {2015})}\BibitemShut {NoStop}%
\bibitem [{\citenamefont {Bernardini}\ \emph {et~al.}(2017)\citenamefont
  {Bernardini}, \citenamefont {Garbarino}, \citenamefont {Sulpice},
  \citenamefont {N{\'u}{\~n}ez-Regueiro}, \citenamefont {Gaudin}, \citenamefont
  {Chevalier}, \citenamefont {Cano},\ and\ \citenamefont
  {Tenc{\'e}}}]{Bernardini2017}%
  \BibitemOpen
  \bibfield  {author} {\bibinfo {author} {\bibfnamefont {F.}~\bibnamefont
  {Bernardini}}, \bibinfo {author} {\bibfnamefont {G.}~\bibnamefont
  {Garbarino}}, \bibinfo {author} {\bibfnamefont {A.}~\bibnamefont {Sulpice}},
  \bibinfo {author} {\bibfnamefont {M.}~\bibnamefont {N{\'u}{\~n}ez-Regueiro}},
  \bibinfo {author} {\bibfnamefont {E.}~\bibnamefont {Gaudin}}, \bibinfo
  {author} {\bibfnamefont {B.}~\bibnamefont {Chevalier}}, \bibinfo {author}
  {\bibfnamefont {A.}~\bibnamefont {Cano}}, \ and\ \bibinfo {author}
  {\bibfnamefont {S.}~\bibnamefont {Tenc{\'e}}},\ }\href
  {http://arxiv.org/abs/1701.05010} {\bibfield  {journal} {\bibinfo  {journal}
  {arXiv:1701.05010 [cond-mat]}\ } (\bibinfo {year} {2017})},\ \bibinfo {note}
  {arXiv: 1701.05010}\BibitemShut {NoStop}%
\bibitem [{\citenamefont {Tence}\ \emph {et~al.}(2009)\citenamefont {Tence},
  \citenamefont {Andre}, \citenamefont {Guadin}, \citenamefont {Bonville},
  \citenamefont {Al~Alam}, \citenamefont {Matar}, \citenamefont {Hermes},
  \citenamefont {Pottgen},\ and\ \citenamefont {Chevalier}}]{Tence2009}%
  \BibitemOpen
  \bibfield  {author} {\bibinfo {author} {\bibfnamefont {S.}~\bibnamefont
  {Tence}}, \bibinfo {author} {\bibfnamefont {G.}~\bibnamefont {Andre}},
  \bibinfo {author} {\bibfnamefont {E.}~\bibnamefont {Guadin}}, \bibinfo
  {author} {\bibfnamefont {P.}~\bibnamefont {Bonville}}, \bibinfo {author}
  {\bibfnamefont {A.~F.}\ \bibnamefont {Al~Alam}}, \bibinfo {author}
  {\bibfnamefont {S.~F.}\ \bibnamefont {Matar}}, \bibinfo {author}
  {\bibfnamefont {W.}~\bibnamefont {Hermes}}, \bibinfo {author} {\bibfnamefont
  {R.}~\bibnamefont {Pottgen}}, \ and\ \bibinfo {author} {\bibfnamefont
  {B.}~\bibnamefont {Chevalier}},\ }\href@noop {} {\bibfield  {journal}
  {\bibinfo  {journal} {J. Applied Physics}\ }\textbf {\bibinfo {volume}
  {106}},\ \bibinfo {pages} {033910} (\bibinfo {year} {2009})}\BibitemShut
  {NoStop}%
\bibitem [{\citenamefont {Tence}\ \emph {et~al.}(2010)\citenamefont {Tence},
  \citenamefont {Matar}, \citenamefont {Andre}, \citenamefont {Gaudin},\ and\
  \citenamefont {Chevalier}}]{Tence2010}%
  \BibitemOpen
  \bibfield  {author} {\bibinfo {author} {\bibfnamefont {S.}~\bibnamefont
  {Tence}}, \bibinfo {author} {\bibfnamefont {S.~F.}\ \bibnamefont {Matar}},
  \bibinfo {author} {\bibfnamefont {G.}~\bibnamefont {Andre}}, \bibinfo
  {author} {\bibfnamefont {E.}~\bibnamefont {Gaudin}}, \ and\ \bibinfo {author}
  {\bibfnamefont {B.}~\bibnamefont {Chevalier}},\ }\href@noop {} {\bibfield
  {journal} {\bibinfo  {journal} {Inorg. Chem.}\ }\textbf {\bibinfo {volume}
  {49}},\ \bibinfo {pages} {4836} (\bibinfo {year} {2010})}\BibitemShut
  {NoStop}%
\bibitem [{\citenamefont {Welter}\ \emph {et~al.}(1998)\citenamefont {Welter},
  \citenamefont {Ijjaali}, \citenamefont {Venturini},\ and\ \citenamefont
  {Malaman}}]{Welter1998}%
  \BibitemOpen
  \bibfield  {author} {\bibinfo {author} {\bibfnamefont {R.}~\bibnamefont
  {Welter}}, \bibinfo {author} {\bibfnamefont {I.}~\bibnamefont {Ijjaali}},
  \bibinfo {author} {\bibfnamefont {G.}~\bibnamefont {Venturini}}, \ and\
  \bibinfo {author} {\bibfnamefont {B.}~\bibnamefont {Malaman}},\ }\href
  {\doibase 10.1016/S0925-8388(97)00280-6} {\bibfield  {journal} {\bibinfo
  {journal} {Journal of Alloys and Compounds}\ }\textbf {\bibinfo {volume}
  {265}},\ \bibinfo {pages} {196} (\bibinfo {year} {1998})}\BibitemShut
  {NoStop}%
\bibitem [{\citenamefont {Giannozzi}\ \emph {et~al.}(2009)\citenamefont
  {Giannozzi}, \citenamefont {Baroni}, \citenamefont {Bonini}, \citenamefont
  {Calandra}, \citenamefont {Car}, \citenamefont {Cavazzoni}, \citenamefont
  {Ceresoli}, \citenamefont {Chiarotti}, \citenamefont {Cococcioni},
  \citenamefont {Dabo}, \citenamefont {Dal~Corso}, \citenamefont
  {de~Gironcoli}, \citenamefont {Fabris}, \citenamefont {Fratesi},
  \citenamefont {Gebauer}, \citenamefont {Gerstmann}, \citenamefont
  {Gougoussis}, \citenamefont {Kokalj}, \citenamefont {Lazzeri}, \citenamefont
  {Martin-Samos}, \citenamefont {Marzari}, \citenamefont {Mauri}, \citenamefont
  {Mazzarello}, \citenamefont {Paolini}, \citenamefont {Pasquarello},
  \citenamefont {Paulatto}, \citenamefont {Sbraccia}, \citenamefont {Scandolo},
  \citenamefont {Sclauzero}, \citenamefont {Seitsonen}, \citenamefont
  {Smogunov}, \citenamefont {Umari},\ and\ \citenamefont
  {Wentzcovitch}}]{Giannozzi2009}%
  \BibitemOpen
  \bibfield  {author} {\bibinfo {author} {\bibfnamefont {P.}~\bibnamefont
  {Giannozzi}}, \bibinfo {author} {\bibfnamefont {S.}~\bibnamefont {Baroni}},
  \bibinfo {author} {\bibfnamefont {N.}~\bibnamefont {Bonini}}, \bibinfo
  {author} {\bibfnamefont {M.}~\bibnamefont {Calandra}}, \bibinfo {author}
  {\bibfnamefont {R.}~\bibnamefont {Car}}, \bibinfo {author} {\bibfnamefont
  {C.}~\bibnamefont {Cavazzoni}}, \bibinfo {author} {\bibfnamefont
  {D.}~\bibnamefont {Ceresoli}}, \bibinfo {author} {\bibfnamefont {G.~L.}\
  \bibnamefont {Chiarotti}}, \bibinfo {author} {\bibfnamefont {M.}~\bibnamefont
  {Cococcioni}}, \bibinfo {author} {\bibfnamefont {I.}~\bibnamefont {Dabo}},
  \bibinfo {author} {\bibfnamefont {A.}~\bibnamefont {Dal~Corso}}, \bibinfo
  {author} {\bibfnamefont {S.}~\bibnamefont {de~Gironcoli}}, \bibinfo {author}
  {\bibfnamefont {S.}~\bibnamefont {Fabris}}, \bibinfo {author} {\bibfnamefont
  {G.}~\bibnamefont {Fratesi}}, \bibinfo {author} {\bibfnamefont
  {R.}~\bibnamefont {Gebauer}}, \bibinfo {author} {\bibfnamefont
  {U.}~\bibnamefont {Gerstmann}}, \bibinfo {author} {\bibfnamefont
  {C.}~\bibnamefont {Gougoussis}}, \bibinfo {author} {\bibfnamefont
  {A.}~\bibnamefont {Kokalj}}, \bibinfo {author} {\bibfnamefont
  {M.}~\bibnamefont {Lazzeri}}, \bibinfo {author} {\bibfnamefont
  {L.}~\bibnamefont {Martin-Samos}}, \bibinfo {author} {\bibfnamefont
  {N.}~\bibnamefont {Marzari}}, \bibinfo {author} {\bibfnamefont
  {F.}~\bibnamefont {Mauri}}, \bibinfo {author} {\bibfnamefont
  {R.}~\bibnamefont {Mazzarello}}, \bibinfo {author} {\bibfnamefont
  {S.}~\bibnamefont {Paolini}}, \bibinfo {author} {\bibfnamefont
  {A.}~\bibnamefont {Pasquarello}}, \bibinfo {author} {\bibfnamefont
  {L.}~\bibnamefont {Paulatto}}, \bibinfo {author} {\bibfnamefont
  {C.}~\bibnamefont {Sbraccia}}, \bibinfo {author} {\bibfnamefont
  {S.}~\bibnamefont {Scandolo}}, \bibinfo {author} {\bibfnamefont
  {G.}~\bibnamefont {Sclauzero}}, \bibinfo {author} {\bibfnamefont {A.~P.}\
  \bibnamefont {Seitsonen}}, \bibinfo {author} {\bibfnamefont {A.}~\bibnamefont
  {Smogunov}}, \bibinfo {author} {\bibfnamefont {P.}~\bibnamefont {Umari}}, \
  and\ \bibinfo {author} {\bibfnamefont {R.~M.}\ \bibnamefont {Wentzcovitch}},\
  }\href {\doibase 10.1088/0953-8984/21/39/395502} {\bibfield  {journal}
  {\bibinfo  {journal} {J. Phys.: Condens. Matter}\ }\textbf {\bibinfo {volume}
  {21}},\ \bibinfo {pages} {395502} (\bibinfo {year} {2009})}\BibitemShut
  {NoStop}%
\bibitem [{\citenamefont {Garrity}\ \emph {et~al.}(2014)\citenamefont
  {Garrity}, \citenamefont {Bennett}, \citenamefont {Rabe},\ and\ \citenamefont
  {Vanderbilt}}]{Garrity2014}%
  \BibitemOpen
  \bibfield  {author} {\bibinfo {author} {\bibfnamefont {K.~F.}\ \bibnamefont
  {Garrity}}, \bibinfo {author} {\bibfnamefont {J.~W.}\ \bibnamefont
  {Bennett}}, \bibinfo {author} {\bibfnamefont {K.~M.}\ \bibnamefont {Rabe}}, \
  and\ \bibinfo {author} {\bibfnamefont {D.}~\bibnamefont {Vanderbilt}},\
  }\href {\doibase 10.1016/j.commatsci.2013.08.053} {\bibfield  {journal}
  {\bibinfo  {journal} {Computational Materials Science}\ }\textbf {\bibinfo
  {volume} {81}},\ \bibinfo {pages} {446} (\bibinfo {year} {2014})}\BibitemShut
  {NoStop}%
\bibitem [{\citenamefont {Perdew}\ \emph {et~al.}(1996)\citenamefont {Perdew},
  \citenamefont {Burke},\ and\ \citenamefont {Ernzerhof}}]{Perdew1996}%
  \BibitemOpen
  \bibfield  {author} {\bibinfo {author} {\bibfnamefont {J.~P.}\ \bibnamefont
  {Perdew}}, \bibinfo {author} {\bibfnamefont {K.}~\bibnamefont {Burke}}, \
  and\ \bibinfo {author} {\bibfnamefont {M.}~\bibnamefont {Ernzerhof}},\ }\href
  {\doibase 10.1103/PhysRevLett.77.3865} {\bibfield  {journal} {\bibinfo
  {journal} {Phys. Rev. Lett.}\ }\textbf {\bibinfo {volume} {77}},\ \bibinfo
  {pages} {3865} (\bibinfo {year} {1996})}\BibitemShut {NoStop}%
\bibitem [{\citenamefont {Moon}\ and\ \citenamefont
  {Choi}(2010)}]{chalcogen_height}%
  \BibitemOpen
  \bibfield  {author} {\bibinfo {author} {\bibfnamefont {C.~Y.}\ \bibnamefont
  {Moon}}\ and\ \bibinfo {author} {\bibfnamefont {H.~J.}\ \bibnamefont
  {Choi}},\ }\href@noop {} {\bibfield  {journal} {\bibinfo  {journal} {Phys.
  Rev. Lett.}\ }\textbf {\bibinfo {volume} {104}},\ \bibinfo {pages} {057003}
  (\bibinfo {year} {2010})}\BibitemShut {NoStop}%
\bibitem [{\citenamefont {Yildirim}(2009)}]{yildirim_physica}%
  \BibitemOpen
  \bibfield  {author} {\bibinfo {author} {\bibfnamefont {T.}~\bibnamefont
  {Yildirim}},\ }\href@noop {} {\bibfield  {journal} {\bibinfo  {journal}
  {Physica C: Superconductivity}\ }\textbf {\bibinfo {volume} {469}},\ \bibinfo
  {pages} {425} (\bibinfo {year} {2009})}\BibitemShut {NoStop}%
\bibitem [{\citenamefont {Yildirim}(2008)}]{Yildirim2008}%
  \BibitemOpen
  \bibfield  {author} {\bibinfo {author} {\bibfnamefont {T.}~\bibnamefont
  {Yildirim}},\ }\href {\doibase 10.1103/PhysRevLett.101.057010} {\bibfield
  {journal} {\bibinfo  {journal} {Phys. Rev. Lett.}\ }\textbf {\bibinfo
  {volume} {101}},\ \bibinfo {pages} {057010} (\bibinfo {year}
  {2008})}\BibitemShut {NoStop}%
\bibitem [{\citenamefont {McMillan}(1968)}]{McMillan1968}%
  \BibitemOpen
  \bibfield  {author} {\bibinfo {author} {\bibfnamefont {W.~L.}\ \bibnamefont
  {McMillan}},\ }\href {\doibase 10.1103/PhysRev.167.331} {\bibfield  {journal}
  {\bibinfo  {journal} {Phys. Rev.}\ }\textbf {\bibinfo {volume} {167}},\
  \bibinfo {pages} {331} (\bibinfo {year} {1968})}\BibitemShut {NoStop}%
\bibitem [{\citenamefont {Boeri}\ \emph {et~al.}(2010)\citenamefont {Boeri},
  \citenamefont {M}, \citenamefont {Mazin}, \citenamefont {Dolgov},\ and\
  \citenamefont {Mauri}}]{boeri2010}%
  \BibitemOpen
  \bibfield  {author} {\bibinfo {author} {\bibfnamefont {L.}~\bibnamefont
  {Boeri}}, \bibinfo {author} {\bibfnamefont {C.}~\bibnamefont {M}}, \bibinfo
  {author} {\bibfnamefont {I.~I.}\ \bibnamefont {Mazin}}, \bibinfo {author}
  {\bibfnamefont {O.~V.}\ \bibnamefont {Dolgov}}, \ and\ \bibinfo {author}
  {\bibfnamefont {F.}~\bibnamefont {Mauri}},\ }\href@noop {} {\bibfield
  {journal} {\bibinfo  {journal} {Phys. Rev. B}\ }\textbf {\bibinfo {volume}
  {82}},\ \bibinfo {pages} {020506} (\bibinfo {year} {2010})}\BibitemShut
  {NoStop}%
\end{thebibliography}%

\end{document}